\documentclass[12pt,preprint]{aastex}
\input psfig.sty
\usepackage{epsf}

\def\vec{\mathbf}
\def\ra{\rangle}
\def\la{\langle}

\def\mpc{\,h^{-1}{\rm Mpc}}

\def\rr{\rho ({\vec r})}
\def\dr{\delta ({\vec r})}

\def\kalias{\vec k+2k_N\vec n}

 1
 
 1

 1
\font\chapter=cmbx10 scaled\magstep 2
 5
 3
\def\gs{\mathrel{\raise1.16pt\hbox{$>$}\kern-7.0pt
\lower3.06pt\hbox{{$\scriptstyle \sim$}}}}
\def\ls{\mathrel{\raise1.16pt\hbox{$<$}\kern-7.0pt
\lower3.06pt\hbox{{$\scriptstyle \sim$}}}}
\def\gtsima{$\; \buildrel > \over \sim \;$}
\def\ltsima{$\; \buildrel < \over \sim \;$}
\def\prosima{$\; \buildrel \propto \over \sim \;$}
\def\gsim{\lower.5ex\hbox{\gtsima}}
\def\lsim{\lower.5ex\hbox{\ltsima}}
\def\simgt{\lower.5ex\hbox{\gtsima}}
\def\simlt{\lower.5ex\hbox{\ltsima}}
\def\simpr{\lower.5ex\hbox{\prosima}}

\begin{document}
\title{Correcting for the alias effect when measuring the power spectrum
using FFT} 
\author {Y.P. Jing} 
\affil {Shanghai Astronomical
Observatory, the Partner Group of MPI f\"ur Astrophysik, \\Nandan Road
80, Shanghai 200030, China}

\begin{abstract}
Because of mass assignment onto grid points in the measurement of the
power spectrum using the Fast Fourier Transform (FFT), the raw power
spectrum $\la |\delta^f(k)|^2\ra$ estimated with FFT is not the same
as the true power spectrum $P(k)$.  In this paper, we derive the
formula which relates $\la |\delta^f(k)|^2\ra$ to $P(k)$. For a sample
of $N$ discrete objects, the formula reads: $\la
|\delta^f(k)|^2\ra=\sum_{\vec n}
[|W(\kalias)|^2P(\kalias)+1/N|W(\kalias)|^2]$, where $W(\vec k)$ is
the Fourier transform of the mass assignment function $W(\vec r)$,
$k_N$ is the Nyquist wavenumber, and $\vec n$ is an integer
vector. The formula is different from that in some of previous works
where the summation over $\vec n$ is neglected. For the NGP, CIC and
TSC assignment functions, we show that the shot noise term $\sum_{\vec
n} 1/N|W(\kalias)|^2]$ can be expressed by simple analytical
functions.  To reconstruct $P(k)$ from the alias sum $\sum_{\vec
n}|W(\kalias)|^2 P(\kalias)$, we propose an iterative method. We test
the method by applying it to an N-body simulation sample, and show
that the method can successfully recover $P(k)$. The discussion is
further generalized to samples with observational selection effects.

\end{abstract}

\keywords {methods: data analysis - methods: statistical - galaxies:
clustering - large-scale structure of Universe }

\section { INTRODUCTION}
The power spectrum $P(k)$ of the spatial cosmic density distribution
is an important quantity in galaxy formation theories. $P(k)$ on large
scales is a direct measure of the primordial density fluctuation, and
$P(k)$ on small scales carries information of later non-linear
evolution, therefore measuring $P(k)$ can serve to distinguish between
different theoretical models.

The power spectrum $P(k)$ as a clustering measure has already been
applied by many authors to observational samples of galaxies and
of clusters of galaxies, including the CfA and Perseus-Pisces redshift
surveys (Baumgart \& Fry 1991), the radio galaxy survey (Peacock \&
Nicholson 1991), the IRAS QDOT survey (Kaiser 1991), the galaxy
distribution in the nearby superclusters (Gramann \& Einasto 1992),
the SSRS redshift sample (Park, Gott, \& da Costa 1992), the CfA and
SSRS extensions (Vogeley et al. 1992; da Costa et al. 1994), the 2Jy
IRAS survey (Jing \& Valdarnini 1993), the 1.2Jy IRAS survey (Fisher
et al. 1993) and the redshift samples of Abell clusters (Jing \&
Valdarnini 1993; Peacock \& West 1992).  The power spectrum is also
widely measured for cosmological N-body simulations, since it can
easily characterize the linear and non-linear evolutions of the
density perturbation (e.g. Davis et al.  1985).

All of these workers except Fisher et al., have used the Fast Fourier
Transformation (FFT) technique to make the Fourier Transforms (FT). In
fact, one can use the direct summation [Eq.~(5) below] to measure the
power spectrum for a sample of a few thousand objects (Fisher et
al. 1993).  However it appears to be more convenient for most workers
to use FFT packages (now available on many computers) to analyze the
power spectrum. This is probably the reason why most of the previous
statistical studies of the power spectrum have used FFT. For N-body
simulations, one has to use FFT to obtain the power spectrum, since
normally there are more than a million particles.  When using FFT, one
needs to collect density values $\rho(\vec r_g)$ on a grid from a
density field $\rho(\vec r)$ (or a particle distribution), which is
usually called `mass assignment'.  The mass assignment is equivalent
to convolving the density field by a given assignment function $W(\vec
r)$ and sampling the convolved density field on a finite number of
grid points. The FT of $\rho(\vec r_g)$ generally is not equal to the
FT of $\rho(\vec r)$, and the power spectrum estimated directly from
the FT of $\rho(\vec r_g)$ is a biased one. The smoothing effect has
already been considered in many previous works (e.g. Baumgart \& Fry
1991; Jing \& Valdarnini 1993; Scoccimarro et al. 1998), but the
sampling effect has not.

In this paper, we derive the formulae which express these effects of
the mass assignment on the estimated power spectrum. A procedure is
proposed to correct for these effects, in order to recover the true
power spectrum. The procedure is tested and shown to be very
successful with a simulation sample.

\section{Formulae}
Let us first recall the definition of the power spectrum $P(k)$.
Let $\rr$ be the cosmic density field and $\overline\rho$ the
mean density. The density field can be 
expressed with a dimensionless field $\dr$ (which is
usually called the density contrast):
\begin{equation}
\dr={\rr-\overline\rho \over \overline\rho}.
\end{equation}
Based on the cosmological principle, one can imagine that 
$\rho(\vec r)$ is periodic in 
some large rectangular volume $V_\mu$. The FT of $\delta(\vec r)$ is 
then defined as:
\begin{equation}
\delta ({\vec k})={1\over V_\mu}\int_{V_\mu}\dr e^{i{\vec r}\cdot{\vec k}}d{\vec r}, 
\end{equation}
and the power spectrum $P(k)$ is simply related to ${\delta ({\vec k})}$ by
\begin{equation}
P(k)\equiv \la\mid{\delta ({\vec k})}\mid^2\ra \,, 
\end{equation}
where $\la\cdot\cdot\cdot\ra $ means the ensemble average.

In practical measurement of $P(k)$ either for an
extragalactic catalogue or for simulation data, the continuous
density field $\rr$ is sampled by only a finite number $N$ of objects. 
In these cases, one needs to deal with the `discreteness' effect 
arising from the Poisson shot noise. 
To show this, let's consider an ideal case where no selection
effect has been introduced into the sample (In fact, some kind of 
selection effects must exist in extragalactic catalogues. We will
discuss it in Section 3). The number density distribution of
objects can be expressed
as $n(\vec r) =\sum_j \delta^D(\vec r -\vec r_j)$, where $\vec r_j$ 
is the coordinate of object $j$ and $\delta^D(\vec r)$ is the
Dirac-$\delta$ function. In analogy with the continuous case, 
the FT of $n(\vec r)$ is defined as:
\begin{equation}
{\delta^d ({\vec k})}
={1\over V_\mu\overline n}\int_{V_\mu} n(\vec r) 
e^{i{\vec r}\cdot{\vec k}}d{\vec r}-\delta^K_{{\vec k},{\vec 0}}\,,
\end{equation}
where $\overline n$ is the global mean number density,
the superscript $d$ represents the {\it discrete} case of $\rr$, and
$\delta^K$ is the Kronecker delta. Following Peebles (1980, sections 36--41),
we divide the volume $V_\mu$ into infinitesimal elements $\{dV_i\}$ with
$n_i$ objects inside $dV_i$. Then the above equation can be written as:
\begin{equation}
{\delta^d ({\vec k})}={1\over N}\sum_i n_i 
e^{i{\vec r}_i\cdot{\vec k}}-\delta^K_{{\vec k},{\vec 0}}\,,
\end{equation}
where $N$ is $\overline n V_\mu$, the number of objects in $V_\mu$. 
Since $dV_i$ is taken so small that $n_i$ is either 0 or 1, we have
$n_i=n_i^2=n_i^3=\cdot\cdot\cdot$, $\la n_i\ra =\overline n dV_i$ and
$\la n_in_j\ra_{i\neq j}={\overline n}^2dV_i dV_j[1+\la \delta (\vec
r_i)\delta (\vec r_j) \ra]$. We
can find the ensemble average of $\delta^d ({\vec k}_1) 
\delta^{d*} ({\vec k}_2)$:
\begin{eqnarray}
\la \delta^d ({\vec k}_1) \delta^{d*} ({\vec k}_2)\ra &=
&{1\over N^2}\sum_{i,j} \la n_i n_j\ra 
e^{i{\vec r}_i\cdot{\vec k}_1-i{\vec r}_j\cdot{\vec k}_2}
-\delta^K_{{\vec k}_1,{\vec 0}}\delta^K_{{\vec k}_2,{\vec 0}}\cr
&=&{1\over N^2}\sum_{i\neq j} \la n_i n_j\ra 
e^{i{\vec r}_i\cdot{\vec k}_1-i{\vec r}_j\cdot{\vec k}_2}+{1\over N^2}\sum_{i} \la n_i\ra 
e^{i{\vec r}_i\cdot({\vec k}_1-{\vec k}_2)} 
-\delta^K_{{\vec k}_1,{\vec 0}}\delta^K_{{\vec k}_2,{\vec 0}}\cr
&=&\la \delta ({\vec k}_1) \delta^* ({\vec k}_2)\ra +{1\over N}
\delta^K_{{\vec k}_1,{\vec k}_2}\,.
\end{eqnarray}
The last equation assumes ${\vec k}_1\neq 0$ or ${\vec k}_2\neq 0$.
The true power spectrum is then:
\begin{equation}
P(k)\equiv \la\mid{\delta ({\vec k})}\mid^2\ra \
=\la \mid{\delta^d ({\vec k})}\mid^2\ra -{1\over N}\,.
\end{equation}
So the discreteness (or shot noise) effect is to introduce an
additional term $1/N$ to the power spectrum $\la \mid{\delta^d ({\vec
k})}\mid^2\ra$.  This fact is already well known to cosmologists. We
present the above derivation because this method is useful in the
following derivations.

In principle one can use the direct summation of Eq.~(5) to measure
the power spectrum for a sample of discrete objects. However, as
described in Section~1, most of the previous statistical studies of
the power spectrum have used FFT. Moreover it would be impossible to
use direct summation to measure the power spectrum for an N-body
simulation.  The quantity computed by the FFT is
\begin{equation}
\delta^{f} ({\vec k})={1\over N}\sum_{\vec g} n^f({\vec r}_g)
e^{i{\vec r}_g\cdot{\vec k}}-\delta^K_{{\vec k},{\vec 0}}\,,
\end{equation}
where the superscript $f$ denotes quantities in FFT. 
$n^f({\vec r}_g)$ is the convolved density value on 
the $\vec g$-th grid point $\vec r_g
=\vec g H$ ($\vec g$ is an integer vector; $H$ is the grid spacing):
\begin{equation}
n^f({\vec r}_g)=\int n({\vec r})W(\vec r-{\vec r}_g) d{\vec r}\,,
\end{equation}
where $W(\vec r)$ is the mass assignment function.

Following Hockney and Eastwood (1981), Eq.(8) can be expressed 
in a more compact way by using the so-called `Sampling Function'. The sampling 
function $\Pi(\vec r)$ is defined as a sum of the 
Dirac-$\delta$ functions spaced 
at unit length in all three spatial directions, i.e. 
$\Pi(\vec r)=\sum_{\{\vec n\}} \delta^D(\vec r-\vec n)$, 
where $\vec n$ is an integer vector.
Defining
\begin{equation}
n'^f(\vec r)\equiv \Pi({\vec r\over H})\int n({\vec r_1})W(\vec r_1-
{\vec r}) d{\vec r_1}
\end{equation}
and constructing
\begin{equation}
\delta'^f ({\vec k})={1\over N}\int n'^f(\vec r) 
e^{i{\vec r}\cdot{\vec k}}d{\vec r}-\delta^K_{{\vec k},{\vec 0}}\,,
\end{equation}
one can easily prove
\begin{equation}
\delta'^f ({\vec k})=\delta^f ({\vec k})\,.
\end{equation}

Therefore one can express $\delta^f ({\vec k})$ as [cf. Eqs.~(10-12)]:
\begin{equation}
\delta^f ({\vec k})={1\over N}\int_{V_\mu} \Pi({\vec r\over H})
\sum_i n_i W({\vec r}_i-{\vec r})
e^{i{\vec r}\cdot{\vec k}}d\vec r-\delta^K_{{\vec k},{\vec 0}}.
\end{equation}
The ensemble average of $\delta^{f} ({\vec k}_1) \delta^{f*} ({\vec k}_2)$
then reads:
\begin{eqnarray}
\la \delta^{f} ({\vec k}_1) \delta^{f*}({\vec k}_2)\ra &=&
{1\over N^2}\int_{V_\mu} \Pi({\vec r_1\over H})\Pi({\vec r_2\over H})
\Bigl[ \sum_{i\neq j} \la n_i n_j\ra W(\vec r_{i1})W(\vec r_{j2})
+\sum_{i}\la n_i\ra W(\vec r_{i1})W(\vec r_{i2})\Bigr] 
\cr
&&\times e^{i{\vec r}_{1}\cdot{\vec k}_1-i{\vec r}_{2}\cdot{\vec k}_2}
d\vec r_1 d\vec r_2
%
%
-{1\over N}\int_{V_\mu} \Pi({\vec r_1\over H})\sum_{i} 
\la n_i\ra W(\vec r_{i1})
e^{i{\vec r_1}\cdot{\vec k}_1}\delta^K_{{\vec k}_2,{\vec 0}}d\vec r_1\cr
&&-{1\over N}\int_{V_\mu} \Pi({\vec r_2\over H})\sum_{i} 
\la n_i\ra W(\vec r_{i2})
e^{-i{\vec r_2}\cdot{\vec k}_2}\delta^K_{{\vec k}_1,{\vec 0}}d\vec r_2
+\delta^K_{{\vec k}_1,{\vec 0}}\delta^K_{{\vec k}_2,{\vec 0}},
\end{eqnarray}
where $\vec r_{ij}=\vec r_i-\vec r_j$. Using
\begin{equation}
\Pi ({\vec k})={1\over V_{\mu}}\int_{V_\mu} \Pi({\vec r\over H}) 
e^{i{\vec r}\cdot{\vec k}}d{\vec r}=\sum_{\vec n}\delta_{\vec k,
{2k_N\vec n}}\,
\end{equation}
where $k_N=\pi/H$ is the Nyquist wavenumber, one can find:
\begin{equation}
\la \delta^{f} ({\vec k}_1) \delta^{f*} ({\vec k}_2)\ra =
\sum_{\vec n_1, \vec n_2} \Bigl[|W({\vec k'}_1)|^2P({\vec k'}_1)
\delta^K_{{\vec k'}_1,{\vec k'}_2}+{1\over N}|W({\vec k'}_1)|^2\delta^K_{{\vec k'}_1,{\vec k'}_2}
\Bigr]\,,
\end{equation}
where $\vec k_i'=\vec k_i+2k_N\vec n_i$ and $W(\vec k)$ is the FT of
$W(\vec r)$. 
For $\vec k_1=\vec k_2=\vec k$ we obtain our desired result:
\begin{equation}
\la |\delta^{f} ({\vec k})|^2\ra = 
\sum_{\vec n} |W(\kalias)|^2P(\kalias)+{1\over N}\sum_{\vec n}|W(\kalias)|^2\,,
\end{equation}
where the summation is over all 3D inetger vectors $\vec n$.  The
meaning of the above equation is very clear. The density convolution
introduces the factor $W^2(\vec k)$ both to the power spectrum and to
the shot noise ($1/N$). The finite sampling of the convolved density
field results in the the `alias' sums (i.e., the sums over $\vec
n$). The alias effect is well-known in Fourier theory, but has not
been taken seriously in the power spectrum analysis of large-scale
clustering in observational cosmology. Both effects of the convolution
and the alias are significant near the Nyquist wavenumber $k_N$, see
Figs. 1 \& 2 in the next section.

\section {  A PROCEDURE TO RECOVER $P(K)$}
In the practical measurement of $P(k)$ using FFT, one should first
choose a mass assignment function. The NGP ($p=1$), CIC ($p=2$) and TSC ($p=3$)
assignment functions are the most popular function for this
purpose. For these schemes, we have
(Hockney \& Eastwood 1981):
\begin{equation}
W(\vec k)=\Bigl[{\sin({\pi k_1\over 2k_N})\sin({\pi k_2\over 2k_N})
\sin({\pi k_3\over 2k_N}) \over ({\pi k_1\over 2k_N})({\pi k_2\over
2k_N})({\pi k_3\over 2k_N})}
\Bigr]^p\,,
\end{equation}
where $k_i$ is the $i$-th component of $\vec k$.

Once one has selected the assignment function, the shot-noise effect
[the second term on the {\it rhs} of Eq.(17)] is easy to correct. For
the NGP, CIC, and TSC assignments, the shot noise term can be
expressed by:
\begin{equation}
D^2(\vec k)\equiv {1\over N}\sum_{\vec n}W^2(\kalias)
={1\over N}C_1(\vec k)\,,
\end{equation}
where $C_1(\vec k)$ are simple analytical functions:
\begin{equation}
C_1(\vec k)=\cases {1, &NGP;\cr
\Pi_{i}[1-{2\over 3}\sin^2({\pi k_i\over 2k_N})], &CIC;\cr
\Pi_{i}[1-\sin^2({\pi k_i\over 2k_N})+
{2\over 15}\sin^4({\pi k_i\over 2k_N})],
&TSC.\cr}
\end{equation}
Furthermore  can easily find out that the $C_1(\vec k)$ of the CIC and
TSC schemes are approximately isotropic for $k\le k_N$, i.e.
\begin{equation}
C_1(\vec k)\approx\cases {[1-{2\over 3}\sin^2({\pi k\over 2k_N})], 
&CIC;\cr
[1-\sin^2({\pi k\over 2k_N})+{2\over 15}\sin^4({\pi k\over 2k_N})],
&TSC.\cr}
\end{equation}

We have tested Equations (19-21) by calculating $\la \mid{\delta^f
({\vec k})}\mid^2\ra $ for 10 random simulation samples, each of which
consists of $N=10^5$ points randomly distributed in a unit cube. In
this case, $D^2(\vec k)= \la \mid{\delta^f ({\vec k})}\mid^2\ra $.  In
the upper panel of Figure 1, we show the average values and $1\sigma$
errors of $\la D^2(\vec k)N/C_1(\vec k)\ra_d$ estimated from the 10
samples, where (and below) $\la\cdot\cdot\cdot\ra_d$ means an average
over all directions of $\vec k$. In this calculation, we have used the
NGP, CIC and TSC assignment functions, and have used Eq.~(20) for
$C_1(\vec k)$.  The result for each assignment is shown by one symbol
in the figure. From Eq.~(19), one expects $\la D^2(\vec k)N/C_1(\vec
k)\ra_d=1$ for all the three assignment functions. Clearly the
simulation results agree very well with Eq.~(20).  In the lower panel,
we show the results in a slightly different way, i.e. the averages and
$1\sigma$ errors of $\la D^2(\vec k)N\ra_d$.  The solid line is $\la
D^2(\vec k)N\ra_d=1$ for the NGP assignment. The dotted and dashed
lines are the approximate expressions of Eq.~(21) for the CIC and TSC
assignments. Again we find a very good agreement between Eq.~(21) and
the results from the random sample. This means that in most
applications, one can use Eq.~(21) to correct the shot noise in the
FFT measurement of the power spectrum.

After correcting the shot noise, our central problem becomes how to
extract $P(k)$ from the first term of Eq.~(17). Let us consider
the correction factor $C_2(k)$ which is defined as 
\begin{equation}
C_2(k)= {\Bigl\la \sum_{\vec n} W^2(\kalias)
P(\kalias)\Bigr\ra_d\over P(k)}\,.
\end{equation}
Since $W(\vec k)$ is a decreasing function,
and $P(k)$ is also a decreasing function 
on the scales $k\ga k_N$, we expect that
the alias contribution of large $\mid \vec n\mid$ to $C_2(k)$ 
is small. In particular, for $k\ll k_N$, any alias contribution is
small and we have $C_2(k)\approx W^2(k)\approx 1$ indpendent of 
$P(k)$. For $k\sim k_N$, the alias contribution to $C_2(k)$ 
becomes important, most of which is owing to the alias 
of $|\kalias| \sim k$. Therefore the dependence of $C_2(k)$ on $P(k)$ 
is only the shape of $P(k)$ at $k\sim k_N$, i.e. the local slope
$\alpha_N$ of $P(k)$ at $k\sim k_N$, $\alpha_N=[\ln P(k)/\ln k]_{k\sim
k_N}$.

Since the local slope $\alpha_N$ is unknown {\it a priori} in
practical measurement of $P(k)$, we propose an iterative method to get
the correction factor $C_2(k)$. Suppose that we have measured the
power spectrum $P_r(k)$:
\begin{eqnarray}
P_r(k)
&=& \Bigl\la\la |\delta^{f} (\vec k)|^2\ra-D^2(\vec k)\Bigr\ra_d\cr
&=& \Bigl\la \sum_{\vec n} W^2(\kalias)P(\kalias)\Bigr\ra_d
\end{eqnarray}
for $k\le k_N$. The local slope $\alpha_0$ of 
$P_r(k)$ at $k\sim k_N$ is calculated by a power-law fitting to 
$P_r(k)$ at $0.5k_N\le k \le k_N$. Assuming
a power-law form $k^{\alpha_0}$ for $P(k)$ of Eq.~(20), we calculate
$C_2(k,\alpha_0)$ and get $P_0(k)=P_r(k)/C_2(k, \alpha_0)$.
Using the local slope $\alpha_0$ at $0.5k_N\le k \le k_N$ of the
$P_0(k)$ just obtained,
we calculate $C_2(k,\alpha_0)$ and $P_0(k)=P_r(k)/C_2(k,
\alpha_0)$ again. This calculation is repeated until $P_0(k)$ (or
$\alpha_0$) converges to some defined accuracy.
The converged $P_0(k)$ is our desired $P(k)$.

Success of the above iteration procedure is shown by Figure 2, where
we present our measurement of $P(k)$ for a simulation sample of Jing
et al. (1995).  The sample is five realizations of P$^3$M simulation
of a low-density flat universe with $\Omega_0=0.2$, $\Lambda_0=0.8$
and $h=1$. The simulation box size is $128\mpc$ and the number of
simulation particles is $N=64^3$. Since the details of the simulation
are unimportant for our result here, we will not discuss them
anymore. The upper panel of the figure shows the raw power spectra
$\la |\delta^{f} (\vec k)|^2\ra$, which are calculated with the NGP,
CIC or TSC mass assignment and with $256^3$ or $64^3$ grid points.  As
expected, the raw power spectrum depends on the scheme of the mass
assignment: higher order mass assignment gives a smaller $\la
|\delta^{f} (\vec k)|^2\ra$ near the Nyquist frequency. At $k\sim
k_N^{64}$, the Nyquist wavenumber of $64^3$ grid points, the raw power
spectra calculated with $256^3$ grid points are expected be very close
to the true power spectrum $P(k)$ because neither effect of the shot
noise [$D^2(k)\ll P(k)$], the convolution [$W(k)\approx 1$] or the
alias [$C_2(k)\approx 1$] is important.  The differences at $k\sim
k_N^{64}$ between the power spectra calculated with $64^3$ grid points
and those with $256^3$ grid points, show the importance of the effects
discussed in this paper.  The lower panel of the figure plots the
power spectra $P(k)$ which are corrected following the procedure
prescribed previously in this section. In this example, we use
Eq.~(21) to correct the shot noise, and we require that the slope
$\alpha_0$ converge to the accuracy of $|\Delta \alpha_0|\le 0.02$ in
the iterative method of deconvolving the alias summation.  For this
accuracy, only fewer than 5 iterations are needed in each
calculation. The six power spectra $P(k)$, measured with different
mass assignments and with different numbers of grid points, agree so
well that their curves overlay each other in the figure. The biggest
difference between the $64^3$ and $256^3$ $P(k)$ is at $k=k_N^{64}$,
which is less than 4 percent.  The result is very encouraging, and it
tells us that using the correction procedure prescribed in this paper,
one can obtain the true $P(k)$ for $k\le k_N$ in the FFT measurement,
independent of which mass assignment is used.

In the above derivations and discussions, we have assumed, for
simplicity, that the sample is uniformly (within Poisson fluctuation)
constructed in a cubic volume. However, the above method is also 
valid for a sample with selection effects (e.g., an extragalactic 
catalogue). To show this, let us introduce 
a selection function $S(\vec r)$ which is defined as the observable rate of
the sample at position $\vec r$. 
If the underlying density distribution is $n(\vec r)$,
the density distribution $n_s(\vec r)$ of the sample is then:
\begin{equation}
n_s(\vec r)=S(\vec r)n(\vec r)\,.
\end{equation}
In this case, we define the following transformation for FFT:
\begin{equation}
\delta_s^{f} ({\vec k})={1\over N}\sum_g [n_s^f({\vec r}_g)-
\overline n S({\vec r}_g)]
e^{i{\vec r}_g\cdot{\vec k}}\,,
\end{equation}
where $n_s^f({\vec r}_g)$ is the convolved density of $n_s(\vec r)$
at ${\vec r}_g$ [cf. Eq.~(9)], and $\overline n$ 
is the mean underlying number density. Following the
derivation of Eq.~(16), we can easily find:
\begin{eqnarray}
\la \delta_s^{f} ({\vec k}_1) \delta_s^{f*} ({\vec k}_2)\ra &=&
\sum_{\vec n_1, \vec n_2} \Bigl[W({\vec k'}_1)W(-{\vec k'}_2)\sum_{{\vec k'}_3}S({\vec k'}_1+{\vec k'}_3)
P({\vec k'}_3)S(-{\vec k'}_2-{\vec k'}_3)\cr
&&+{1\over N}S({\vec k'}_1-{\vec k'}_2)W({\vec k'}_1)W(-{\vec k'}_2)
\Bigr]\,,
\end{eqnarray}
where $S(\vec k)$ is defined as:
\begin{equation}
S (\vec k)={1\over \int S(\vec r)d\vec r}\int S(\vec r) 
e^{i{\vec r}\cdot{\vec k}}d{\vec r}.
\end{equation}
Because $S(\vec k)$ peaks at $\vec k\approx 0$, the coupling of the 
selection and the power spectrum  [the first term of Eq.~(26)] is 
important only at small $k$. How to treat this coupling is a nontrivial
task and beyond the scope of this paper (but see e.g. Peacock \& Nicholson
1991; Jing \& Valdarnini 1993). However for larger $k$ where the 
effects discussed here become significant, we have:
\begin{eqnarray}
\la |\delta^{f} ({\vec k})|^2\ra \approx
\sum_{\vec k_3}S^2(\vec k_3)\sum_{\vec n} W^2(\kalias)P(\kalias)
+{1\over N}\sum_{\vec n}W^2(\kalias)
\end{eqnarray}
The difference between Eq.~(26) and Eq.~(19) is only the factor
$\sum_{\vec k_3}S^2(\vec k_3)$. Therefore our procedure for correcting
the effects of the mass assignment is still valid for observational
samples with selection effects.  Equation (28) is derived on the
assumption that $S(\vec k)$ is compact in $\vec k$ space. This
assumption is valid for many redshift surveys, e.g., the CfA and IRAS
redshift surveys. But is not valid for surveys of irregular
boundaries, e.g., pencil beam surveys. In addition, current redshift
surveys of galaxies already cover a volume of $(1000\mpc)^3$. Because
one is also interested in the clustering information down to scales of
$0.1\mpc$, a $(10,000)^3$ gridpoint FFT is required to explore the
clustering on all scales, which is still difficult to be realized on
modern supercomputers. Other estimators, e.g., the FT of the two-point
correlation function (Jing \& B\"orner 2004), should be used to
determine $P(k)$ on small scales.

\section {SUMMARY}
In this paper, we have derived for the first time the formula 
[Eq.~(17)] which relates the raw power spectrum $\la |\delta^f
(k)|^2\ra$ estimated with FFT to the true power spectrum 
$P(k)$. The formula shows clearly how the mass assignment modifies
the power spectrum. The convolution of the density field with an assignment 
function $W(\vec r)$ reshapes the power spectrum  (including the shot
noise spectrum) by multiplying the factor $|W(\vec k)|^2$; the finite sampling
of the convolved density field leads to the alias sum. 
We have described how to reconstruct $P(k)$ from $\la |\delta^f
(k)|^2\ra$. For the NGP, CIC and TSC assignment functions, the shot
noise $D^2(\vec k)$ can be expressed by simple analytical functions,
therefore the shot noise can be easily corrected. To extract $P(k)$
from the alias sum $\sum_{\vec n}W^2(\kalias)P(\kalias)$, we propose 
an iterative method. The method has been tested by applying it to
an N-body simulation sample. Using different numbers of grid points, we
have shown that the method can very successfully recover $P(k)$ for
all the NGP, CIC and TSC assignment functions.

\acknowledgements

The author thanks Bhuvnesh Jain for stimulating discussion on an
earlier version of the draft. The work is supported in part by NKBRSF
(G19990754) and by NSFC (Nos.10125314, 10373012).

\begin{figure}
\epsscale{1.0} \plotone{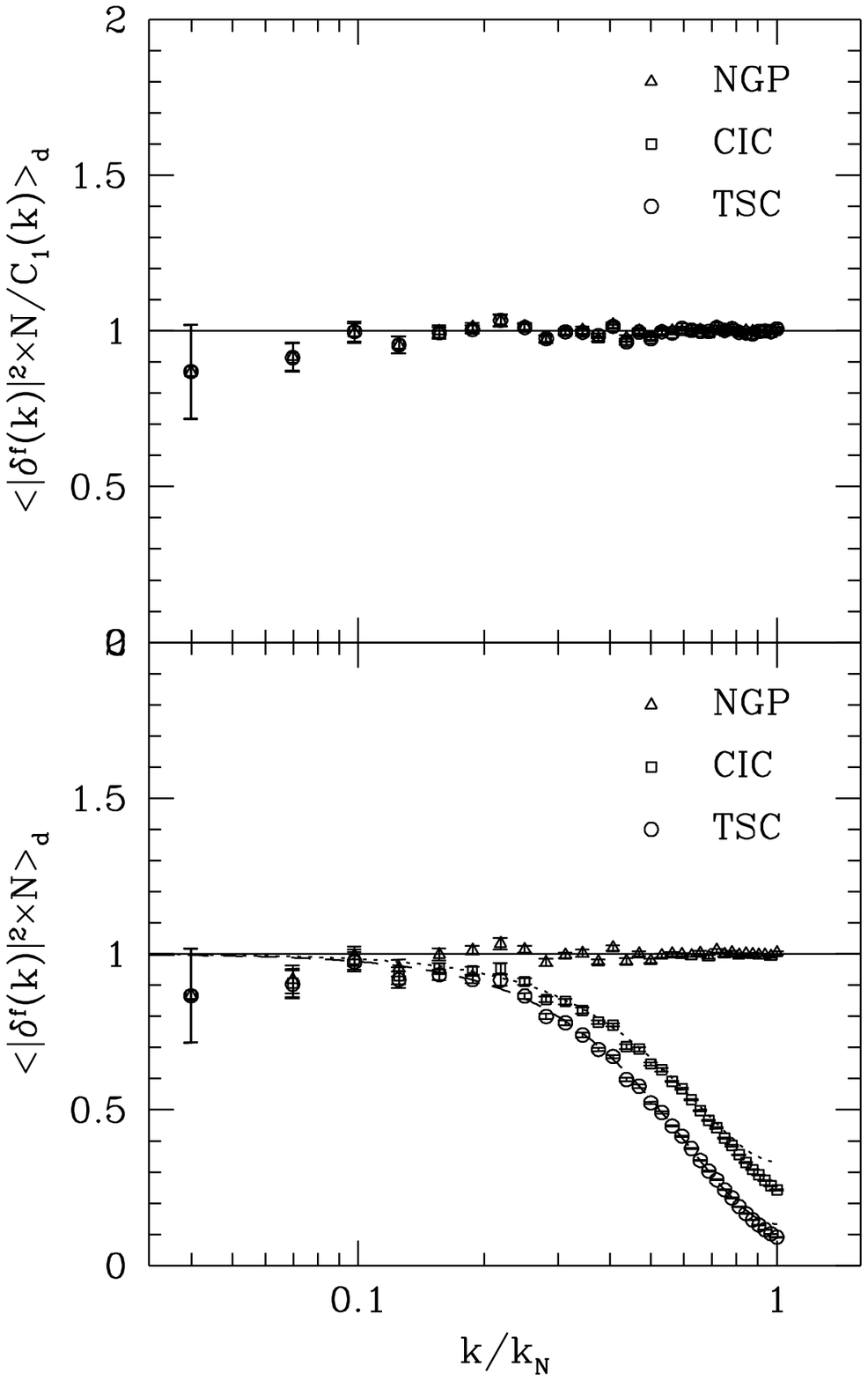}
\caption{The shot noise $D^2(\vec k)$ estimated from 10 samples of
Poisson distributed random points. Each symbol represents the result
for each mass assignment, as indicated in the figure.  The upper panel
shows the result with the function $\la D^2(\vec k)N/C_3(\vec
k)\ra_d$, i.e. the $D^2(\vec k)$ scaled to $1/N C_1(\vec k)$. The
estimated result agrees quite well with the analytical prediction $\la
D^2(\vec k)N/C_3(\vec k)\ra_d=1$.  The lower panel compares the
estimated $\la D^2(\vec k)N\ra_d$ with our analytical predictions for
the NGP (solid line), CIC (dotted line) and TSC (dashed line)
assignment functions. For CIC and TSC, we have used the approximate
formulae of Eq.~(21).}
\end{figure}

\begin{figure}
\epsscale{1.0} \plotone{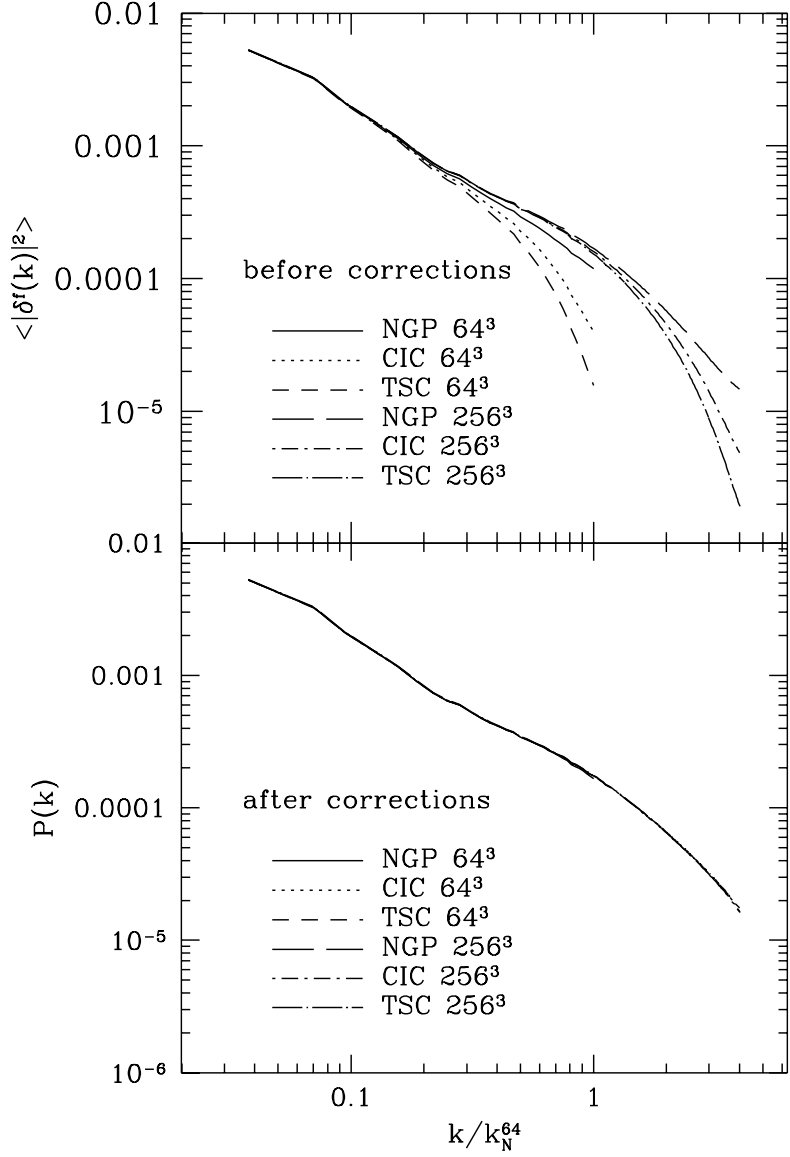}
\caption {The six power spectra which we measured for an N-body
simulation sample using three mass assignments (NGP, CIC and TSC) and
two grids ($64^3$ and $256^3$ grid points). $k_N^{64}$ is the Nyquist
wavenumber of $64^3$ grid points. The upper panel shows the raw power
spectra $\la|\delta^f(k)|^2\ra$ estimated directly from the FFT
(Eq.~8). The lower panel shows the true power spectra $P(k)$
reconstructed from the $\la|\delta^f(k)|^2\ra$ following the procedure
described in the text. The six reconstructed $P(k)$ agree so well that
their curves overlay each other.}
\end{figure}

\end{document}